\documentclass[aps,pre,10pt,superscriptaddress,balancelastpage,notitlepage,twocolumn]{revtex4-2}
\usepackage[english]{babel}
\usepackage[utf8]{inputenc}
\usepackage[colorinlistoftodos, color=green!40, prependcaption]{todonotes}
\usepackage{caption}
\usepackage{centernot}
\usepackage{graphicx}
\usepackage{amsmath}
\usepackage{times}
\usepackage{amssymb}
\usepackage{mathrsfs}
\usepackage{color}
\usepackage{url}
\usepackage{version}
\usepackage[pdftex,colorlinks=true,pdfstartview=FitV,linkcolor= linkcolor,citecolor= linkcolor,urlcolor= linkcolor,hyperindex=true,hyperfigures=false]{hyperref}
\usepackage{balance}
\usepackage{lastpage}
\usepackage{parskip}
\usepackage[paper=a4paper,margin=1in]{geometry}
  
\definecolor{linkcolor}{rgb}{0,0,0.6} 

\newcommand{\vect}[1]{\boldsymbol{#1}}

\begin{document}

\title{An Introduction to Motility-Induced Phase Separation}

\author{Joakim Stenhammar}
\email{joakim.stenhammar@fkem1.lu.se}
\affiliation{Division of Physical Chemistry, Lund University, P.O. Box 124, S-221 00 Lund, Sweden}

\begin{abstract}
\noindent Motility-induced phase separation, MIPS, is arguably the most well-studied collective phenomenon occurring in active matter without alignment interactions (\emph{scalar} active matter). Its basic origin is simple: since self-propelled particles accumulate where they move slowly, having a propulsion speed that decreases steeply enough with density, due to collisions or chemical interactions, leads to a feedback loop that induces the formation of a dense phase. In these notes, I will discuss some of the main theoretical and computational efforts that have been made over the last decade in understanding the basic structural and dynamical properties of MIPS phase coexistence in microscopic active particle models. 
\end{abstract}

\maketitle

\section{Introduction} 
\noindent The phenomenon of \emph{motility-induced phase separation} (MIPS) is one of the most well-studied collective behaviours in models (and, to some extent, experimental realisations) of active matter. As the name indicates, MIPS involves a phase coexistence between two phases of different densities, similar to what happens in a binary fluid mixture as the temperature is decreased below the critical temperature. The difference between MIPS and a traditional phase separation is that the driving force is completely non-thermodynamic: while phase separation in equilibrium fluids is driven by attractive interparticle interactions that overcome the entropy of mixing as quantified by the thermal energy $k_B T$ (a \emph{static} property), MIPS is instead induced by the non-equilibrium character of the single-particle motion (a \emph{dynamic} property). MIPS has turned out to be a robust phenomenon occurring in a range of active matter models without alignment interactions~\citep{Cates:2015:ARCMP,Tiribocchi:CRPhys:2015} (\emph{scalar} active matter) in contact with a solid substrate or other momentum sink (\emph{dry} active matter). Due to its conceptual similarity to equilibrium phase separation, much effort has been made to understand MIPS using tools and concepts from classical (equilibrium) thermodynamics~\citep{Palacci:PRL:2010,Tailleur:2008:PRL,Wilding:PRL:2021,Solon:PRE:2018,Schmidt:PRE:2019,Paliwal:NJP:2018,Mallory:PRE:2014,Takatori:PRE:2015,Pagonabarraga:PRL:2018,Klamser:NatCommun:2018}, recast to take into account the non-equilibrium character of the system. These notes will address some of these developments, focusing on computational and theoretical descriptions of MIPS in microscopic active particle models. Due the very extensive literature on the topic, the aim is not to give a comprehensive review of the results in the field, but rather focus on a few key concepts. We will first illustrate the mechanistic origin of pattern formation and MIPS in active particles that interact via a position- or density-dependent propulsion speed, before switching to active particles that interact through pairwise forces. In particular, we will address the concept of pressure in suspensions of active Brownian particles, and to what extent this can be used as a ``thermodynamic'' variable to describe phase coexistence. Finally, we will briefly discuss some of the open research problems within the MIPS field. 

\begin{figure}[h]
    \centering
    \includegraphics[width=7.0cm, clip, trim=0.22cm 0.12cm 0.05cm 0.05cm]{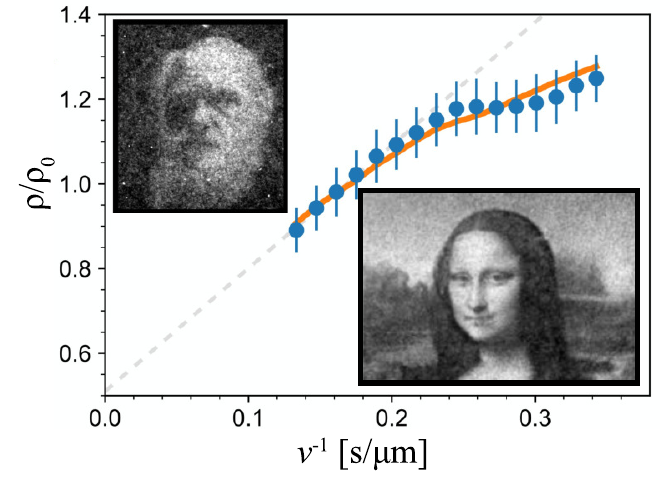}
    \caption{\textbf{Accumulation of swimming \emph{E. coli} bacteria using spatially varying light intensities.} The plot shows the measured local density as a function of the swimming speed, with the predicted $\rho \sim v^{-1}$ behaviour shown by the dashed line. The deviations at low speeds is primarily due to interactions between bacteria that limit their local density. The insets show ``density-painted'' pictures of Mona Lisa and Charles Darwin, achieved by adaptively varying the local light intensity. Adapted from~\citep{diLeonardo:elife:2018}.}\label{Monalisa}
\end{figure}

\section{Pattern formation in active particles with spatially varying speeds}
\noindent Arguably, the simplest example of an active system is a single self-propelled particle that moves with constant speed $v_0$ in a direction $\vect{n}$ that relaxes either continuously through rotational Brownian motion (active Brownian particles) or via discontinuous tumbling events (run-and-tumble particles). This directed motion requires either the consumption of chemical fuel, or some sort of external driving (\emph{e.g.}, by light), both of which drive the particle out of thermodynamic equilibrium. Now, we divide our physical system into two compartments where the particle has two distinct swimming speeds $v_1$ and $v_2$, where $v_2 > v_1$. This setup can be most easily imagined for the case of a particle driven by an external light field~\citep{Palacci:Science:2013,diLeonardo:elife:2018,Arlt:SciAdv:2018} of varying intensity, but can also be imagined to arise through a nonuniform distribution of nutrients in a suspension of swimming bacteria. Everything else alike, it is then obvious that the active particle will spend more time in the region where it moves slower, so that, over long times, the particle density $\rho$ obeys $\rho_1 > \rho_2$; in other words, active particles accumulate where they move slowly. This is easily imagined in everyday settings -- think about traffic congestion near a roadwork area with reduced speed -- but it is important to remember that this is an intrinsically non-equilibrium phenomenon: in equilibrium, the mean-square velocity $\langle v^2 \rangle$ of a particle is related to the temperature $T$ by the equipartition theorem $\langle v^2 \rangle = 3 k_B T / m$. Thus, the only way to create a nonuniform velocity for a Brownian particle is to have a nonuniform temperature, destroying the equilibrium nature of the system. 

To analyse this accumulation effect in quantitative terms, we follow~\citet{Schnitzer:PRE:1993} and consider run-and-tumble particles in $d=1$ dimension: this system is composed of right- and left-moving particles that move in a motility landscape $v(x)$ and tumble with spatially varying frequency $\alpha(x)$. As you will show in Exercise 1, the particle current $J(x)$ in this system can be expressed as:
\begin{equation}\label{eq:QSAP_flux}
J(x) = -\frac{v^2(x)}{\alpha(x)} \frac{d \rho}{dx} - \frac{v(x) \rho(x)}{\alpha(x)} \frac{dv}{dx}.
\end{equation}
Assuming the existence of a flux-free steady state ($J=0$), this equation can be solved for the density distribution $\rho(x)$, yielding~\citep{Schnitzer:PRE:1993,Cates:2015:ARCMP}
\begin{equation}\label{eq:vrho_const}
\rho(x) = \rho_0 \frac{v_0}{v(x)},
\end{equation}
where $\rho_0$ is the reference density in a system with particle speed $v_0$. For our specific case of two different swim speeds, this means that $\rho_1 v_1 = \rho_2 v_2$, \emph{i.e.}, the local particle density is inversely proportional to the local particle speed but independent of the tumbling rate. This simple yet powerful result has been elegantly demonstrated in experiments on light-powered swimming bacteria~\citep{diLeonardo:elife:2018,Arlt:SciAdv:2018}: if a properly shaped ``light pattern'' of varying intensity is used, such bacteria can be directed to dynamically self-assemble into intricate (and artistic) patterns (see Fig.~\ref{Monalisa}). 

\section{MIPS in quorum-sensing active particles}\label{sec:QSAP}
\noindent We now consider the closely related case where the active particle speed instead depends on position implicitly through a dependence on the local particle density, \emph{i.e.}, $v(x) = v[\rho(x)]$~\citep{Tailleur:2008:PRL,Cates:2015:ARCMP}. From a microbiological view, this constitutes a minimal model of \emph{quorum sensing}, whereby bacteria chemically sense and respond to their local density by a change of gene expression (phenotype). The phenotypic change can constitute deflagellation or a decrease in swimming speed, for example during the initial stages of biofilm formation~\citep{Polin:elife:2021}.  

\begin{figure}[h]
    \centering
    \includegraphics[width=7.5cm, clip, trim=0.0cm 0cm 0.0cm 0.0cm]{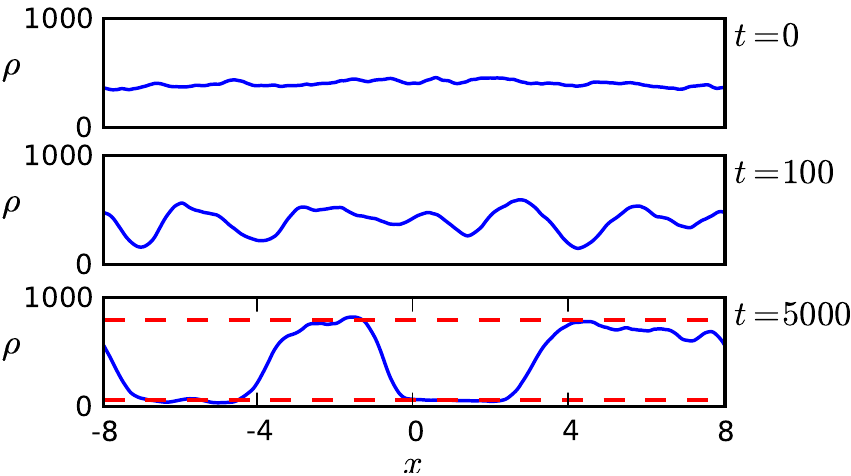}
    \caption{\textbf{MIPS in quorum-sensing particles.} For densities where the instability criterion~\eqref{MIPS_criterion} is fulfilled, an initially homogeneous system ($t=0$) will phase separate into coexisting regions of high and low densities. In 1D, the domains maintain a finite size at steady state ($t=5000$) due to the finite energy cost of forming an interface, while for $d>1$ the domains will coarsen until bulk phase separation is reached. Reproduced with permission from~\citep{Tailleur:2008:PRL} (\copyright 2008, American Physical Society).}\label{QSAPs}
\end{figure}

Following~\citet{Cates:2015:ARCMP}, we consider a small perturbation $\delta \rho$ around a homogeneous state of particle density $\rho_0 \equiv c/v(\rho_0)$ with $c$ some constant. The speed at the perturbed density is given by 
\begin{equation}
    v\left[ \rho_0 + \delta \rho(x) \right] = v(\rho_0) + v'(\rho_0) \delta \rho(x),
\end{equation}
where $v'(\rho) \equiv \frac{dv}{d\rho}$. Thus, if $v'< 0$, a positive density perturbation ($\delta \rho > 0$) leads to a velocity decrease ($\delta v < 0$). For steeply enough decreasing $v(\rho)$, this leads to a further increase in the local density to $\rho_0 + \delta \hat{\rho}$, triggering a feedback loop that leads to MIPS. More precisely, to linear order in $\delta \rho$ we have
\begin{equation}\label{eq:delta_rho}
\begin{split}
    \rho_0 + \delta \hat{\rho} & = \frac{c}{v(\rho_0) + v'(\rho_0) \delta \rho} \\ & \approx \rho_0 \left[ 1 - \frac{v'(\rho_0)}{v(\rho_0)}\delta \rho \right].
\end{split}
\end{equation}
The condition for the onset of the MIPS feedback loop is that $\delta \hat{\rho} > \delta \rho$, leading to
\begin{equation}\label{MIPS_criterion}
    \frac{v'(\rho_0)}{v(\rho_0)} < -\frac{1}{\rho_0}.
\end{equation}
This constitutes a linear instability criterion on the homogeneous density $\rho_0$: for $\rho > \rho_0$, the homogeneous state will destabilise, leading to a spinodal decomposition into coexisting dense and dilute regions, as shown in Fig.~\ref{QSAPs} for the choice of an exponentially decreasing $v(\rho)$~\citep{Tailleur:2008:PRL}. Interestingly, as you will show in Excercise 2, the MIPS criterion~\eqref{MIPS_criterion} can be reformulated as an equilibrium spinodal condition $d^2 f_0 / d\rho^2 < 0$ on an effective bulk free energy density $f_0(\rho)$ given by~\citep{Tailleur:2008:PRL} 
\begin{equation}\label{f_0}
    f_0(\rho) = \rho(\ln \rho - 1) + \int^{\rho} \ln[v(u)] du.
\end{equation}
Here, the first term can be identified as the ideal gas entropy, which favours a homogeneous particle distribution, and the second term encodes an effective attraction due to the decreasing $v(\rho)$ responsible for MIPS. This thermodynamic mapping of MIPS to phase coexistence in equilibrium fluids breaks down at higher orders in a gradient expansion of the free energy, \emph{i.e.}, for interfacial free energy terms~\citep{Tailleur:2008:PRL,Stenhammar:2013:PRL,Wittkowski:2014:NC}. Nevertheless, it points towards strong analogies between MIPS and equilibrium phase coexistence at the mesoscopic level, in spite of the strongly nonequilibrium dynamics at the microscopic level. 

\section{Microscopic particle models for MIPS}

\noindent In this Section, we will briefly review the three main particle-based models used for theoretical and computational studies of MIPS, namely quorum-sensing active particles (QSAPs), active Brownian particles (ABPs), and active Ornstein-Uhlenbeck particles (AOUPs). While the qualitative properties of MIPS can be readily demonstrated in 1D (see Fig.~\ref{QSAPs}), studying 2- and 3-dimensional systems is necessary to gain a quantitative understanding of the phase diagram and coarsening dynamics. Since the differences between MIPS in 2D and 3D are small~\citep{Stenhammar:SoftMatter:2014,Wysocki:2014:EPL}, we will restrict ourselves to $d=2$ in the following.

\begin{figure}[h]
    \centering
    \includegraphics[width=6.5cm, clip, trim=0.0cm 0.0cm 0.0cm 0.0cm]{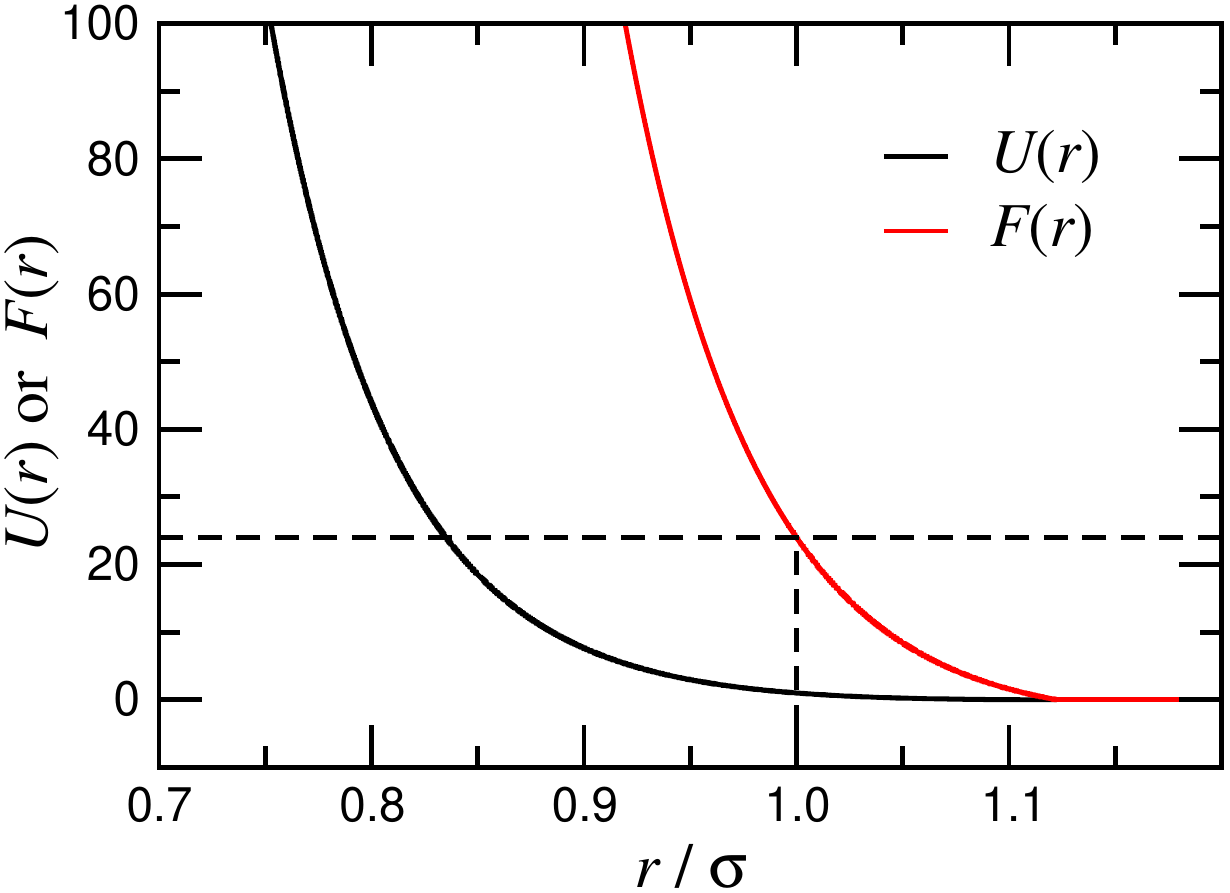}
    \caption{\textbf{WCA potential.} Interparticle potential $U(r)$ and force $F(r) = -dU/dr$ for the WCA potential in Eq.~\eqref{eq:WCA}, expressed in units of $\varepsilon$ and $\varepsilon / \sigma$. Using the constant self-propulsion force $F_p = v_0 = 24$ (dashed line) is common, as it leads to an ``effective diameter'' of $\sim \sigma$ for two colliding particles~\citep{Stenhammar:SoftMatter:2014}.}\label{fig:WCA}
\end{figure}

The 1D, isotropic QSAP model introduced in the previous section is readily extended to $d = 2$, yielding the following equation of motion for the position $\vect{r}_i$ of particle $i$ with orientation $\vect{n}_i$:
\begin{equation}\label{QSAP}
    \partial_t \vect{r}_i = v[\tilde{\rho}(\vect{r}_i)] \vect{n}_i + \sqrt{2D_T} \vect{\Lambda}_T, \quad \mathrm{(QSAPs)}
\end{equation}
where the second term represents Brownian diffusion with translational diffusion coefficient $D_T$ and $\vect{\Lambda}_T$ a zero-mean, unit variance Gaussian white noise process that obeys $\langle \Lambda_{i\alpha}(t) \Lambda_{j \beta}(t') \rangle = \delta_{ij} \delta_{\alpha \beta} \delta(t-t')$. In computational implementations, the choice of functional form for $v(\rho)$ needs to be complemented by a recipe for measuring the local density field $\tilde{\rho}$. This is typically achieved by coarse-graining the instantaneous particle density $\hat{\rho}(\vect{r}) = \sum_i \delta (\vect{r}-\vect{r}_i)$ in the neighbourhood of each particle by convoluting it with an isotropic coarse-graining kernel $K$~\citep{Solon:NJP:2018}:
\begin{equation}\label{eq:rhotilde}
\tilde{\rho}(\vect{r}) = \int K(\vect{r}-\vect{r}') \hat{\rho}(\vect{r}')d\vect{r}'.
\end{equation}
A typical choice for $K$ is the short-ranged, bell-shaped kernel~\citep{Solon:NJP:2018}
\begin{equation}
K(r) = \frac{\Theta(r_0-r)}{Z} \exp\left(-\frac{r_0^2}{r_0^2 - r^2} \right),
\end{equation}
where $\Theta$ is the Heaviside function, $Z$ a normalisation constant and $r_0$ the cutoff radius. The dynamics of the particle orientation $\vect{n}_i$ can furthermore be implemented either as a run-and-tumble process with frequency $\alpha$ as in 1D, or as Brownian rotational diffusion with diffusion constant $D_R$ (which can only be realised for $d \geq 2$), as for ABPs (see Eq.~\eqref{ABP_rot}); from the perspective of MIPS, these two choices are essentially equivalent upon the mapping $D_R \longleftrightarrow 2(d-1)\alpha$~\citep{Cates:EPL:2013}. 

A more computationally straightforward way to treat interparticle interactions is through a constant swimming speed $v_0$ combined with pairwise interparticle forces $\vect{F}_i(\{ \vect{r}_j \})$:
\begin{equation}\label{PFAP}
    \partial_t \vect{r}_i = v_0 \vect{n}_i + \vect{F}_i(\{ \vect{r}_j \}) + \sqrt{2D_T} \vect{\Lambda}_T, \; \mathrm{(ABPs)}.
\end{equation}
Here, we have implicitly set the single-particle mobility to unity, which, together with the overdamped dynamics, enables the implementation of self propulsion as an effective ``self-propulsion force'' $F_p = v_0$. As discussed in previous chapters, for ABPs, Eq.~\eqref{PFAP} is combined with a diffusional relaxation of the swimming direction $\vect{n}_i \equiv (\cos \theta_i, \sin \theta_i)$: 
\begin{equation}\label{ABP_rot}
    \partial_t \theta_i = \sqrt{2D_R} \Lambda_R, \quad \mathrm{(ABPs)}
\end{equation}
where $D_R$ is the rotational diffusion constant and $\Lambda_R$ is a white-noise process. Equations~\eqref{PFAP}--\eqref{ABP_rot} together constitute the active Brownian particle (ABP) model~\citep{Romanczuk:EPJST:2012}, which has become the mainstay model for computational studies of MIPS. 

A third well-studied model model for active particles undergoing MIPS is the active Ornstein-Uhlenbeck particle (AOUP) model~\citep{vanWijland:PRE:2021,Szamel:PRE:2014,Maggi:SciRep:2015,Eichhorn:FrontPhys:2021}, which replaces the constant self-propulsion term with a translational noise with memory (coloured noise), and therefore does not require a separate description of the orientational dynamics. The equation of motion for an AOUP interacting through pairwise forces is given by
\begin{equation}\label{AOUP1}
    \partial_t \vect{r}_i = \vect{v}_i + \vect{F}_i(\{ \vect{r}_j \}), \quad \mathrm{(AOUPs)}
\end{equation}
where the velocity $\vect{v}_i$ is the solution of
\begin{equation}\label{AOUP2}
    \tau \partial_t \vect{v}_i = -\vect{v}_i + \sqrt{2D_A} \vect{\Lambda}_A. \quad \mathrm{(AOUPs)}
\end{equation}
Here, $\tau$ is the persistence time of the propulsion direction, $D_A$ is an ``active diffusion coefficient'' describing the persistent random walk process, and $\vect{\Lambda}_A$ is a white noise with the same properties as $\vect{\Lambda}_T$ in~\eqref{PFAP}. The solution of \eqref{AOUP2} yields an exponentially correlated velocity, $\langle v_{i\alpha} v_{j\beta} \rangle = \delta_{ij} \delta_{\alpha \beta} \frac{D_A}{\tau} e^{-|t|/\tau}$. The AOUP model simplifies analytical progress, while sharing the qualitative features of MIPS in ABPs.~\citep{vanWijland:PRE:2021}

While all three models described above exhibit MIPS, the field-theoretical description of QSAPs is qualitatively different from that of ABPs and AOUPs~\citep{Solon:NJP:2018}, since QSAPs interact in a non-pairwise manner. Nevertheless, as we will see below, the $v(\rho)$ dynamics in QSAPs can to some extent be viewed as a mean-field description of the repulsive pairwise forces in ABPs, allowing for easier analytical progress. 

Due to their popularity, in the following we will mainly focus on MIPS in ABP suspensions. The computational treatment of ABP suspensions is straightforward, and the dynamics of Eqs.~\eqref{PFAP} and~\eqref{ABP_rot} can be solved using standard techniques from molecular dynamics simulations. Since the ABP dynamics are overdamped (\emph{i.e.}, the particle velocity $\partial_t \vect{r}_i(t)$ is directly proportional to the force $\vect{F}_i(t)$), they can be integrated using a simple Euler scheme. The interparticle force $\vect{F}_i(\{ \vect{r}_j \})$ in Eq.~\eqref{ABPs} is typically pairwise additive and isotropic, \emph{i.e.}, $\vect{F}_i = \sum_{j \neq i} \vect{F}_{ij}$, where $\vect{F}_{ij} = -\nabla U(r_{ij})$ is the force derived from the interaction potential $U$ and $r_{ij} = |\vect{r}_i - \vect{r}_j|$. A particularly common choice for $U(r)$ is the short-ranged, repulsive Weeks-Chandler-Andersen (WCA) potential depicted in Fig.~\ref{fig:WCA}, given by
\begin{equation}\label{eq:WCA}
U(r) =  \begin{cases} 
		4\varepsilon \left[ \left( \frac{\sigma}{r}\right)^{12} - \left( \frac{\sigma}{r} \right)^{6} \right] + \varepsilon \; \text{for}\; r < 2^{1/6} \sigma \\
		0 \quad \text{otherwise}
		\end{cases}
\end{equation}
where $\sigma$ is the effective particle diameter, and $\varepsilon$ controls the steepness of the potential. Due to the short range of the WCA potential, the computation time can be made linear in the number of particles $N$ by using standard tools from molecular simulations such as neighbour lists. Thus, simulating systems of $N > 10^4$ ABPs is relatively straightforward on standard computers, while system sizes of $N > 10^7$ have been simulated on supercomputers~\citep{Stenhammar:SoftMatter:2014}, enabling the resolution of mesoscopic structure and dynamics of systems undergoing MIPS in both 2D and 3D. 

\section{MIPS in active Brownian particles}

\begin{figure}[h]
    \centering
    \includegraphics[width=7.6cm, clip, trim=0.0cm 0cm 0.0cm 0.0cm]{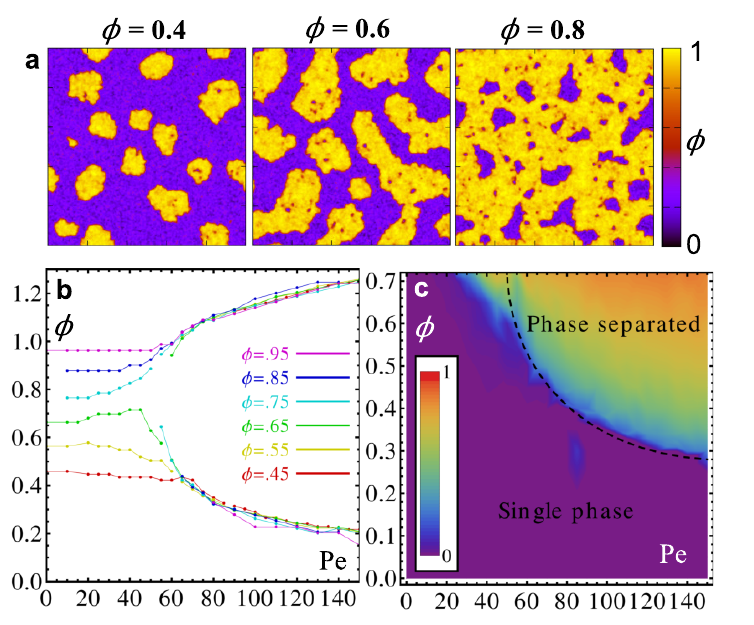}
    \caption{\textbf{Phase coexistence in ABPs.} \textbf{(a)} Snapshots of the local packing fraction $\phi = \rho \pi \sigma^2 / 4$ in ABP suspensions of three different overall densities undergoing MIPS. Reproduced from~\citep{Stenhammar:SoftMatter:2014}. \textbf{(b)} Measured coexistence densities as a function of Pe and $\phi$ for varying overall densities, as indicated. For $\mathrm{Pe} > \mathrm{Pe}_c \approx 60$, the suspension separates into two coexisting phases, whose densities are independent of the overall density. \textbf{(c)} Cluster fraction $\varphi_c$ (Eq.~\eqref{f_c_def}) measured from simulations in the low- and intermediate-density region. The dashed line shows the analytical prediction for $\varphi = 0$ from Eq.~\eqref{f_c} with $q = 4.5$. Panels bc are reproduced with permission from~\citep{Redner:2013:PRL} (\copyright 2013, American Physical Society).}\label{ABPs}
\end{figure}

\noindent MIPS in 2D suspensions of ABPs was first realised in 2012 by~\citet{Fily:2012:PRL}, who observed that, starting from a disordered initial configuration, dense and dilute phases would quickly form for high enough particle swimming speeds and densities. These results were expanded by~\citet{Redner:2013:PRL}, who characterized MIPS through a phase diagram (Fig.~\ref{ABPs}b) in terms of the density and the dimensionless P\'eclet number, defined by
\begin{equation}\label{Peclet}
    \mathrm{Pe} = \frac{3v_0}{D_R \sigma},
\end{equation}
where the factor 3 comes from the connection between the translational and rotational diffusion constants in equilibrium, $D_R = 3 D_T / \sigma^2$. Apart from this numerical factor, Pe measures the persistence length $l_p = v_0 \tau_R $ of the single-particle motion normalised by the particle diameter $\sigma$, where $\tau_R = D_R^{-1}$ is the rotational relaxation time. Pe plays the role of an inverse temperature in determining the phase diagram shown in Fig.~\ref{ABPs}b: For $\mathrm{Pe} > \mathrm{Pe}_c \approx 60$, the homogeneous system phase separates into dense and dilute phases with densities $\rho_\ell$ and $\rho_g$, independent of the overall packing fraction $\rho$. 

\noindent Phenomenologically, the mechanism for MIPS in QSAPs discussed in Section~\ref{sec:QSAP} can be applied also to the case of ABP suspensions: collisions between particles slow them down in areas of high density, leading to a decreasing $v(\rho)$, which causes MIPS for large enough particle densities. For low and intermediate densities, the collision frequency between ABPs is proportional to density, while the average duration of a collision event is density independent, leading to a linearly decreasing effective $v(\rho)$:
\begin{equation}\label{eq:v_rho}
v(\rho) \simeq v_0 (1-\rho/\rho_*),
\end{equation}
where $\rho_*$ is a near close-packed density above which $v = 0$. ABP simulations for $\mathrm{Pe} < \mathrm{Pe}_c$ confirm Eq.~\eqref{eq:v_rho} up to densities very close to $\rho_*$, where $v(\rho)$ goes smoothly to zero and higher-order terms in $\rho$ become important~\citep{Stenhammar:SoftMatter:2014,Solon:NJP:2018} (see Exercise 4). Furthermore, $v(\rho)$ has a negligible dependence on Pe for $\mathrm{Pe} < \mathrm{Pe}_c$. Inserting~\eqref{eq:v_rho} into the instability criterion~\eqref{MIPS_criterion} leads to the simple condition $\rho > \rho_*/2$ for MIPS. While reasonably close to the observed critical density in the high-Pe limit, this condition lacks a dependence on both $v_0$ and $D_R$ (or $\alpha$), in qualitative contrast to the computational observation of a $\mathrm{Pe}_c \gg 1$. This is an effect of the isotropic nature of the coarse-grained density $\tilde{\rho}$ in~\eqref{eq:rhotilde}, as discussed further in the next section. Instead using the coarse-grained density $\tilde{\rho}(\vect{r}_i+\xi \vect{n}_i)$ a distance $\xi > 0$ in front of the particle, we can introduce an anisotropy into the dynamics, and yields a dependence on $v_0$ and $D_R$ that qualitatively captures the Pe dependence in the ABP phase diagram~\citep{Solon:NJP:2018}. Rather than giving a full account of the anisotropic QSAP description, we will in the next section instead review the kinetic route developed by~\citet{Redner:2013:PRL} to explain the strong Pe-dependence of the ABP coexistence densities. 

\subsection{Kinetic model of MIPS in ABPs}

\noindent We start by assuming a dense phase with a (Pe-independent), close-packed density $\rho_\ell = 2/(\sqrt{3}\sigma^2)$ surrounded by an ABP gas of (Pe-dependent) density $\rho_g$. We furthermore assume that the area of the dense cluster is large enough to yield an effectively flat vapour-liquid interface. The incoming flux of particles per unit length from the gas to the liquid is then $k_{\mathrm{in}} \sim \rho_g v_0$. The outgoing flux is incorporated by noticing that a particle will exit the cluster into the gas phase whenever its orientation $\vect{n}$ has diffused above the ``horizon'', \emph{i.e.}, $\vect{n}\cdot \vect{\hat{s}} > 0$, where $\vect{\hat{s}}$ is the outward surface normal~\citep{Redner:2013:PRL}. Since $\vect{n}$ evolves through rotational diffusion, the outward flux depends on $D_R$, yielding $k_{\mathrm{out}} \sim D_R / \sigma$. Since particles are assumed to be completely trapped as long as they are oriented below the horizon, $k_{\mathrm{out}}$ is independent of $v_0$. Crucially, this $D_R$ dependence is absent in the effective $v(\rho)$ picture, since the speed of a QSAP is independent of its orientation as long as $\tilde{\rho}$ in Eq.~\eqref{QSAP} is measured isotropically. 

At steady state, $k_{\mathrm{in}} = k_{\mathrm{out}}$, which leads to an expression for $\rho_g$ in terms of $v_0$ and $D_R$. Taking into account prefactors, this leads to~\citep{Redner:2013:PRL} $\rho_g = \pi q D_R / (v_0 \sigma)$, where the free parameter $q \geq 1$ incorporates the fact that, when a particle escapes the cluster, it will lead to the escape of additional subsurface particles that had previously been trapped by it. Using the expressions for $\rho_g$ and $\rho_\ell$, we can now calculate the fraction $\varphi_c$ of particles in the cluster phase, defined as
\begin{equation}\label{f_c_def}
    \varphi_c \equiv \frac{\rho_\ell V_\ell}{\rho_\ell V_\ell + \rho_g V_g} = \frac{\rho \rho_\ell - \rho_\ell \rho_g}{\rho \rho_\ell - \rho \rho_g},
\end{equation}
where $V_\ell$ and $V_g$ are the respective volumes of the two phases, and $\rho$ is the overall density of the system. Inserting the expressions for $\rho_\ell$ and $\rho_g$ and using the definition~\eqref{Peclet} of Pe yields the following expression for $\varphi_c (\mathrm{Pe}, \rho)$~\citep{Redner:2013:PRL}:
\begin{equation}\label{f_c}
    \varphi_c (\mathrm{Pe}, \rho) = \frac{2 \sigma^2 \rho \mathrm{Pe} - 6 \pi q}{2 \sigma^2 \rho \mathrm{Pe} - 3\sqrt{3} \pi q\sigma^2 \rho}.
\end{equation}
In this kinetic picture, the Pe dependence thus arises as a direct consequence of the balance between the $v_0$-dependent inwards flux and the $D_R$-dependent outwards flux. The line $\varphi_c = 0$ corresponds to the stability of the dense phase in the thermodynamic limit, equivalent to the binodal line, yielding $\rho \sim 1/\mathrm{Pe}$. (This should be contrasted with the QSAP criterion in Eq.~\eqref{MIPS_criterion} which considers the linear stability of the \emph{homogeneous} phase, corresponding to the \emph{spinodal} line). Restoring dimensional quantities, the binodal criterion becomes $v_0 \rho \sigma \gtrsim
D_R$, where the left-hand side can be identified as the average collision frequency in the homogeneous phase. This criterion shows that the reorientation time $\tau_R$, which sets the typical time needed for two colliding particles to separate, needs to be longer than the time between particle collisions for MIPS to occur. In other words, two particles need to stay together long enough after a collision that a third particle can collide with them in order for a cluster to be stable. In Fig.~\ref{ABPs}c, the binodal line $\varphi_c = 0$ from Eq.~\eqref{f_c} is plotted together with $\varphi_c$ measured from ABP simulations~\citep{Redner:2013:PRL}, showing that the kinetic picture sketched above indeed describes the gas binodal at moderate densities. 

\begin{figure}[h]
    \centering
    \includegraphics[width=6.5cm, clip, trim=0.0cm 0.0cm 0.0cm 0.0cm]{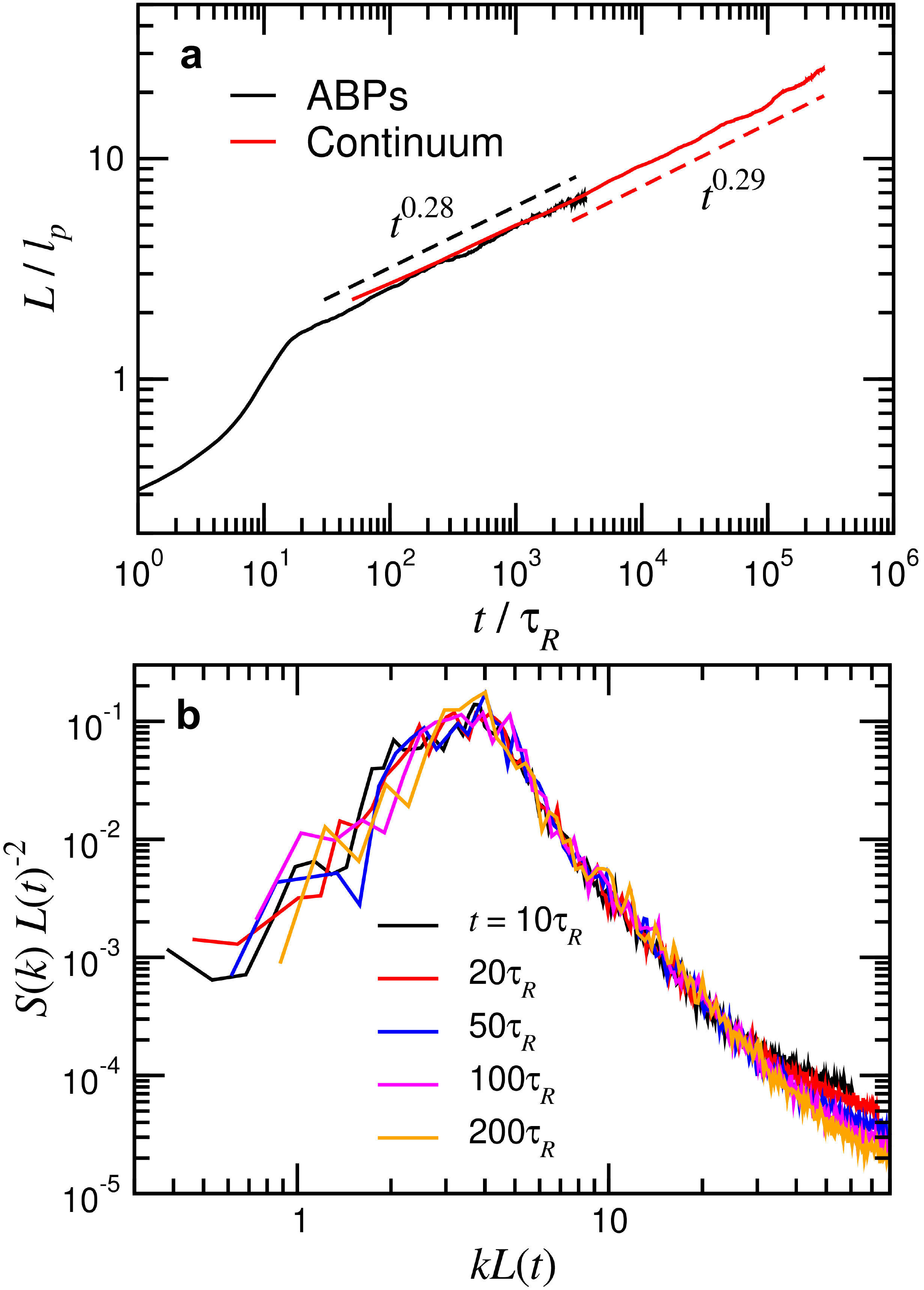}
    \caption{\textbf{Coarsening kinetics of ABPs.} \textbf{(a)} Growth of the characteristic cluster size $L(t)$ with time in an ABP suspension undergoing MIPS (black) and in a field theory based on the effective free energy density of Eq.~\eqref{f_0} with a linearly decaying $v(\rho)$ measured from simulations (red). $L(t)$ is normalised by the persistence length $l_p = v_0 \tau_R$. The dashed lines indicate approximate growth exponents; the small deviations from the predicted $t^{1/3}$ scaling is an intermittent effect due to the high noise level. \textbf{(b)} Structure factor $S(k)$ measured at different times during coarsening. When rescaled properly by $L(t)$, the curves collapse onto each other due to self-similarity of the domain morphology. Reproduced from~\citep{Stenhammar:SoftMatter:2014}.} \label{Kinetics}
\end{figure}

\subsection{Similarities between MIPS and equilibrium phase coexistence}

\noindent In spite of its intrinsically non-equilibrium origin, MIPS shares a number of similarities with simple equilibrium fluids phase separating due to attractive interactions, with Pe playing the role of an inverse temperature: \\
\begin{enumerate}
    \item When changing the average density $\rho$ inside the two-phase region, $\rho_g$ and $\rho_\ell$ remain constant (Fig.~\ref{ABPs}ab), while the relative volumes of the two phases change continuously according to the lever rule $N_g/N_\ell = (\rho_\ell - \rho)/(\rho - \rho_g)$, where $N_g$ and $N_\ell$ are the number of particles in the dilute and dense phases, respectively~\citep{Redner:2013:PRL}. \\
        \item The ABP phase diagram has both binodal and spinodal lines (Fig.~\ref{Binodals_Maxwell}b), where the region between them is characterised by a metastable homogeneous phase, which eventually undergoes phase separation through nucleation and growth, while systems near and inside the spinodal region decompose instantaneously~\citep{Solon:PRE:2018,Solon:NJP:2018}. \\
        
    \item The phase separation kinetics are consistent with the scaling law $L(t) \sim t^{1/3}$, where $L(t)$ measures the typical size of phase-separated domains as a function of time (Fig.~\ref{Kinetics}a)~\citep{Stenhammar:2013:PRL,Stenhammar:SoftMatter:2014}. In equilibrium, this scaling is expected in systems where coarsening is driven by surface tension leading to diffusive transport of material from small droplets to large ones, so-called Ostwald ripening. \\
    
    \item The evolving domain pattern after a quench is self-similar, in the sense that the domain patterns, as measured through the structure factor $S(q)$ at different times, are statistically identical when rescaled by $L(t)$ (Fig.~\ref{Kinetics}b)~\citep{Stenhammar:2013:PRL,Stenhammar:SoftMatter:2014}. \\
\end{enumerate}
These observed similarities have inspired a number of studies connecting MIPS to equilibrium thermodynamic concepts such as chemical potential~\citep{Stenhammar:2013:PRL,Tailleur:2008:PRL,Paliwal:NJP:2018}, pressure~\citep{Brady:2014:PRL,Pagonabarraga:PRL:2018,Solon:PRL:2015,Winkler:SoftMatter:2015,Speck:PRE:2016} and temperature~\citep{Mallory:PRE:2014,Ginot:2015:PRX,Petrelli:PRE:2020}, redefined to take into account the intrinsic non-equilibrium character of ABPs. For example, using the QSAP model as a mean-field description for ABPs, the bulk free energy in Eq.~\eqref{f_0} can be combined with a suitable interfacial tension-like term (which in contrast does not correspond to any equilibrium counterpart) to build a continuum model of the phase separation kinetics in MIPS (Fig.~\ref{Kinetics}a). Such field theories accurately reproduce the coarsening behaviour (points 3 and 4 above) in ABPs undergoing MIPS for deep quenches~\citep{Stenhammar:2013:PRL,Stenhammar:SoftMatter:2014}. However, to capture the coexistence densities, it turns out that a description based on pressure in coexisting phases is more fruitful, which we will address in the next section. 

\section{Mechanical pressure in ABPs}\label{sec:pressure}

\begin{figure}[h]
    \centering
    \includegraphics[width=6.5cm, clip, trim=0.0cm 9.45cm 0.0cm 0.0cm]{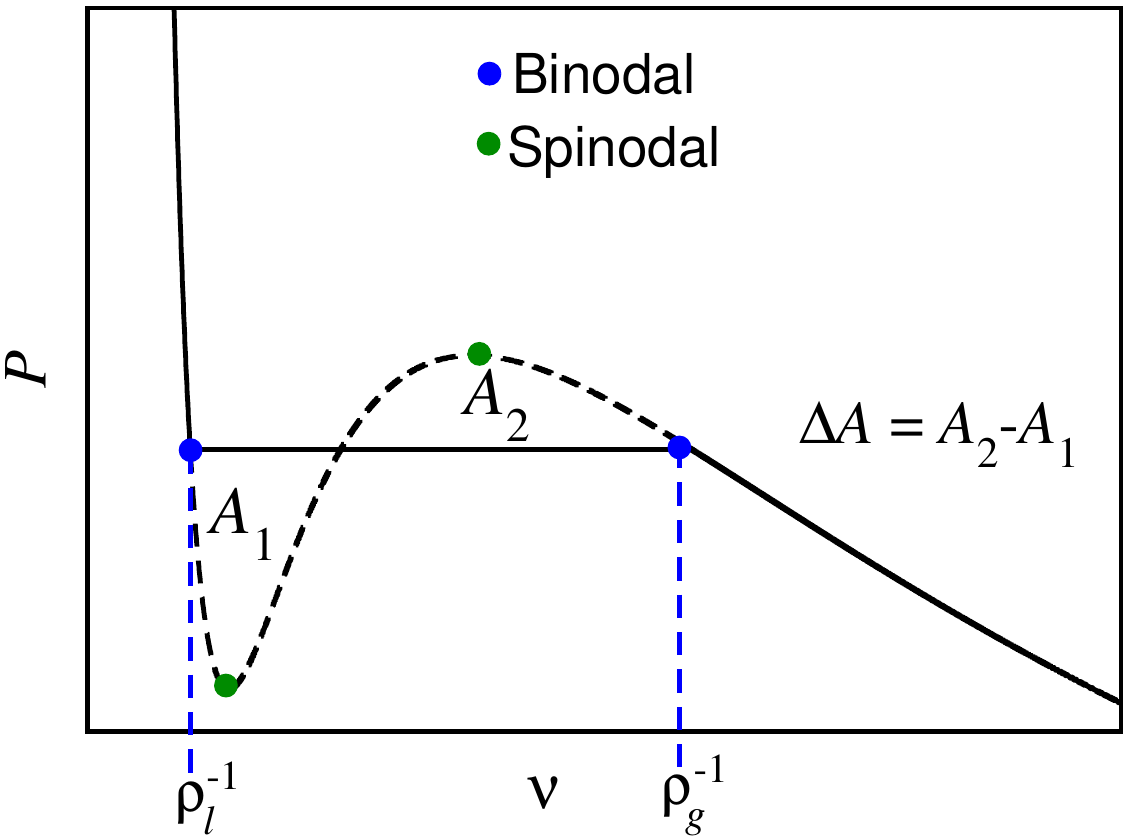}
    \caption{\textbf{Maxwell equal-area construction.} In equilibrium fluids, the coexistence densities $\rho_g$ and $\rho_\ell$ are determined by the equality of pressures and chemical potentials between the two bulk phases. The first condition implies a horizontal line connecting the two phases, while the latter geometrically corresponds to $\Delta A = A_2 - A_1 = 0$, leading to the binodal densities indicated by blue circles. The spinodal region, corresponding to the linear instability of the homogeneous phase, is given by the condition $dP/d\nu \geq 0$ (green circles). In the two-phase region, the system follows the horizontal straight line, corresponding to a constant pressure, while the dashed line corresponds to the underlying (unphysical) EOS of the one-phase system. For ABPs, the pressure remains constant in the two phases, while the equal-area construction on $P$ fails, leading to $\Delta A \neq 0$ (Eq.~\eqref{MC} and Fig.~\ref{Binodals_Maxwell}a).}\label{Maxwell}
\end{figure}

\noindent In a phase-separating equilibrium fluid, the coexistence densities are determined by the equality between the two phases of two bulk quantities: the chemical potential $\mu$ and the pressure $P$, defined respectively as derivatives of the free energy with respect to density and volume. In addition to this thermodynamic definition, the pressure also has a \emph{mechanical} definition, as the force per unit area measured either on the container walls or transmitted across an imaginary plane inside the bulk fluid, the so-called virial pressure. In equilibrium these definitions must be equivalent, and, since the free energy depends on bulk properties only, the mechanical pressure must be independent of the particle-wall potential: the pressure is an \emph{equation of state} (EOS) in equilibrium. The EOS is usually expressed as pressure $P$ as a function of either the density $\rho$ or the volume per particle $\nu \equiv \rho^{-1}$ at constant temperature. As long as $P(\nu)$ is a monotonically decreasing function, the homogeneous system is stable for all compositions. For parameters where $P(\nu)$ is non-monotonic, the system becomes mechanically (and thermodynamically) unstable for some compositions and separates into coexisting phases of densities $\rho_g = \nu_g^{-1}$ and $\rho_\ell = \nu_\ell^{-1}$, whose values can be determined through the ``Maxwell equal-area construction'' on $P(\nu)$~\citep{Callen:Thermodynamics} as demonstrated in Fig.~\ref{Maxwell}. 

\subsection{Pressure in noninteracting ABPs}\label{P_nonint}

\noindent Since ABP suspensions are far from equilibrium, we cannot \emph{a priori} assume them to have any mapping to thermodynamic quantities such as pressure or chemical potential, although the mechanical pressure is still a well-defined quantity. For simplicity, we start by considering the pressure in a suspension of \emph{non-interacting} ABPs, confined by a flat wall acting on the particle $x$-coordinate \emph{via} a force $\vect{F}_w(x) = \vect{\hat{x}} F_w(x)$; since the system is translation invariant along the $y$ direction, we consider only the $x$-dependence of the relevant quantities. Following~\citet{Solon:2015:NP}, we write the master equation for the probability $\psi (x,\theta,t)$ of finding a particle at position $x$ pointing in the direction $\theta$ at time $t$ as
\begin{equation}\label{master_eqn}
\begin{split}
    \partial_t \psi(x,\theta,t) & = -\partial_x \left[ v_0 \cos \theta + F_w(x) \right] \psi \\ & + D_T \partial_x^2 \psi + D_R \partial_\theta^2 \psi.
\end{split}
\end{equation}
We now define the $n^{\mathrm{th}}$ moment of $\psi$ as $m_n = \int_0^{2\pi} \cos(n\theta)\psi(x,\theta) d\theta$, where $m_0 = \rho(x)$ corresponds to the density, $m_1 = \mathcal{P}(x)$ to the polarization, and $m_2 = \mathcal{Q}(x)$ to the nematic order, where the two latter are measured normal to the interface. Taking the first two moments of Eq.~\eqref{master_eqn} (see Exercise 3) and setting $\partial_t \rho = \partial_t \mathcal{P} = 0$, we obtain

\begin{align}
    &  v_0 \mathcal{P}(x) + F_w(x) \rho(x) = D_T \partial_x \rho (x), \label{m0_eqn} \\
    & D_R\mathcal{P}(x) = -\partial_x \Big[ \frac{v_0}{2}(\mathcal{Q}(x) + \rho(x)) \label{m1_eqn} \\ \nonumber
     & + F_w(x) \mathcal{P}(x) - D_T \partial_x \mathcal{P}(x) \Big].
\end{align}

\begin{figure}[h]
    \centering
    \includegraphics[width=6.5cm, clip, trim=0.0cm 0.0cm 0.0cm 0.0cm]{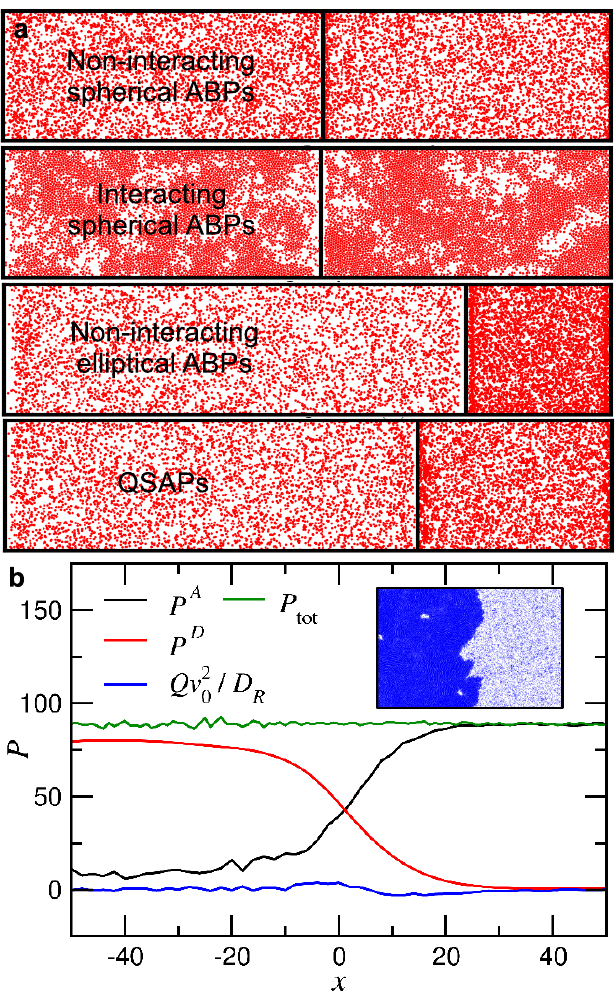}
    \caption{\textbf{Pressure is an equation of state for spherical ABPs.} \textbf{(a)} Snapshots of steady-state configurations in two subsystems separated by a mobile,  asymmetric wall with different $F_w$ on its two sides. For noninteracting ABPs and for ABPs interacting via the spherical WCA potential~\eqref{eq:WCA}, the density, and thus the bulk pressure, is independent of the wall potential, showing that mechanical pressure is an EOS in these systems. For elliptical ABPs that experience particle-wall torques and for QSAPs, the bulk density however depends on the wall potential, showing that pressure is not an EOS. Reproduced with permission from~\citep{Solon:2015:NP} (\copyright 2015, Springer Nature). \textbf{(b)} Components of the pressure measured across a liquid-gas interface (inset). The active pressure $P^A = P_0^A + P_1^A$ and direct pressure $P^D = P_0^D + P_1^D$ contain both bulk and interfacial contributions, while the nematic contribution $\mathcal{Q}v_0^2 / D_R$ is zero away from the interface. The sum $P_{\mathrm{tot}}$ of the three terms is constant across the interface, verifying that mechanical pressure is equal in coexisting phases. Reproduced from~\citep{Solon:NJP:2018}.} \label{Pressure_EOS}
\end{figure}

The mechanical pressure $P$ exerted on a wall at $x \gg 0$ can now be calculated \emph{via} the relation~\citep{Solon:2015:NP}
\begin{equation}\label{P_def}
    P = -\int_0^\infty \rho(x) F_w(x) dx.
\end{equation}
Inserting Eqs.~\eqref{m0_eqn}--\eqref{m1_eqn} and integrating yields the following, remarkably simple, expression for the ``ideal'' pressure $P_I$ of a suspension of noninteracting ABPs at density $\rho$~\citep{Solon:2015:NP,Brady:2014:PRL,Winkler:SoftMatter:2015,Speck:PRE:2016}: 
\begin{equation}\label{P_id}
    P_I(\rho) = \left[ D_T + \frac{v_0^2}{2D_R} \right]\rho.
\end{equation}
Here, we can identify the first term as the ideal gas pressure $\rho k_B T$ (note that $D_T = k_B T$ since we have set the single-particle mobility to unity), while the second term equals $\rho$ times the ``active diffusion constant'' $v_0^2/(2D_R)$. Crucially, this expression is independent of $F_w$: thus, mechanical pressure is an EOS for ABPs, and can potentially be used as an effective thermodynamic quantity (see Fig.~\ref{Pressure_EOS}). As shown in~\citep{Solon:2015:NP}, this property is not true for general self-propelled particle systems, but only as long as the dynamics of $\vect{n}$ are decoupled from the translational dynamics. Any torque between the particles and the wall introduces a dependence on $F_w$ in the expression for $P_I$, and destroys the EOS property of the mechanical pressure, as illustrated in Fig.~\ref{Pressure_EOS}a. This constraint is clearly not present in equilibrium fluids, where any nonspherical molecule will experience a torque close to a surface, yet pressure remains an EOS in such systems. 

\subsection{Pressure of interacting ABPs}

\noindent The case of ABPs interacting both with each other and with the walls can be handled using a similar approach as for noninteracting particles (Section~\ref{P_nonint}), with the master equation~\eqref{master_eqn} extended to include the effects of interparticle interactions. These not only lead to an additional term containing the interparticle force $\vect{F}$, but also to a description in terms of the \emph{fluctuating} hydrodynamic fields $\hat{\psi}$ and $\hat{\rho}$~\citep{Solon:PRL:2015}:
\begin{equation}\label{master_eqn_int}
\begin{split}
    & \partial_t \hat{\psi}(\vect{r}) = \\ & -\nabla \cdot \left[ (v_0\vect{n} + \vect{F}_w)\hat{\psi} + \int \vect{F}(\vect{r}'-\vect{r})\hat{\rho}(\vect{r}')\hat{\psi}(\vect{r}) d\vect{r}' \right] \\ & + D_T \nabla^2 \hat{\psi} + D_R \partial_\theta^2 \hat{\psi} \\ & + \nabla \cdot \left( \sqrt{2 D_T \hat{\psi}} \mathbf{\Lambda} \right) +  \partial_\theta \left( \sqrt{2 D_R \hat{\psi}} \xi \right), 
\end{split}
\end{equation}
where $\mathbf{\Lambda}$ and $\xi$ are $\delta$-correlated, unit-variance noise fields. While the coarse-graining process leading to~\eqref{master_eqn_int} in the interacting case is more demanding due to the multiplicative noise~\citep{Dean:1996:JPA}, a similar overall strategy as in the noninteracting case can be used to derive the following expression for the mechanical pressure $P_0(\rho)$ in a suspension of interacting ABPs at homogeneous density $\rho$~\citep{Solon:PRL:2015,Brady:2014:PRL,Winkler:SoftMatter:2015}:
\begin{equation}\label{P_bulk}
\begin{split}
    P_0 & = D_T \rho + P_0^D + P_0^A \\ 
    P_0^D & \equiv \int_{x>\Lambda} dx \int_{x'<\Lambda} d\vect{r}' F_x(\vect{r}-\vect{r}') \langle \hat{\rho}(\vect{r}) \hat{\rho}(\vect{r}') \rangle \\
    P_0^A & \equiv \frac{v_0 v(\rho)}{2 D_R} \rho,
\end{split}
\end{equation}
where we call $P_0^D$ the \emph{direct pressure} and $P_0^A$ the \emph{active pressure}. The integral in the definition of $P_0^D$ measures the average interparticle force $F_x$ transmitted across a virtual plane $\Lambda$ perpendicular to the $x$ direction in the bulk fluid. This expression is equivalent to that of the standard virial pressure in equilibrium fluids~\citep{Takatori:PRE:2015,Solon:PRL:2015}, and can be more succinctly expressed as $P_0^D = (2V)^{-1} \langle \vect{F}_i \cdot \vect{r}_i \rangle$, with $V$ the total system volume. While equivalent to its equilibrium definition, $P_0^D$ implicitly depends on activity through the correlator $\langle \hat{\rho}(\vect{r}) \hat{\rho}(\vect{r}') \rangle$. $P_0^A$, on the other hand, is an intrinsically active contribution that vanishes in the passive limit $v_0 \rightarrow 0$. It differs from the active term in the ideal pressure~\eqref{P_id} in that one of the $v_0$ factors has been changed into the average, density dependent particle speed $v(\rho)$, measured along the particle orientation vector $\vect{n}_i$, \emph{i.e.}, $v(\rho) \equiv v_0 + \langle \vect{F}_i \cdot \vect{n}_i \rangle$. For purely repulsive forces, $\vect{F}_i$ is typically antiparallel to $\vect{n}_i$, since head-on collisions are more frequent than rear collisions; thus, $v(\rho) \leq v_0$, in accordance with the mean-field QSAPs mapping. Just as for non-interacting ABPs, neither of the terms in~\eqref{P_bulk} depends on $\vect{F}_w$, meaning that the mechanical pressure remains an EOS (Fig.~\ref{Pressure_EOS}a); crucially, however, the EOS property breaks down for QSAPs, which interact through non-pairwise forces~\citep{Solon:2015:NP}. In addition to the requirement of no particle-wall torques, the EOS property for interacting ABPs holds only as long as there are no interparticle alignment interactions. These two requirements are clearly violated in most (if not all) experimental examples of ABPs, where particles will reorient either due to their anisotropic shapes (such as for swimming bacteria) or due to long-ranged hydrodynamic or chemotactic interactions. Nevertheless, analysing pressure as an effective thermodynamic variable in the idealised, torque-free ABP limit can give important fundamental knowledge about the possibilities (and limitations) of equilibrium mappings in active matter systems. In the next section, we will thus connect the EOS property of pressure in interacting ABP suspensions with their MIPS coexistence densities. 

\subsection{Pressure and phase coexistence in ABPs}

\begin{figure}[h]
    \centering
    \includegraphics[width=6.5cm, clip, trim=0.0cm 0.0cm 0.0cm 0.0cm]{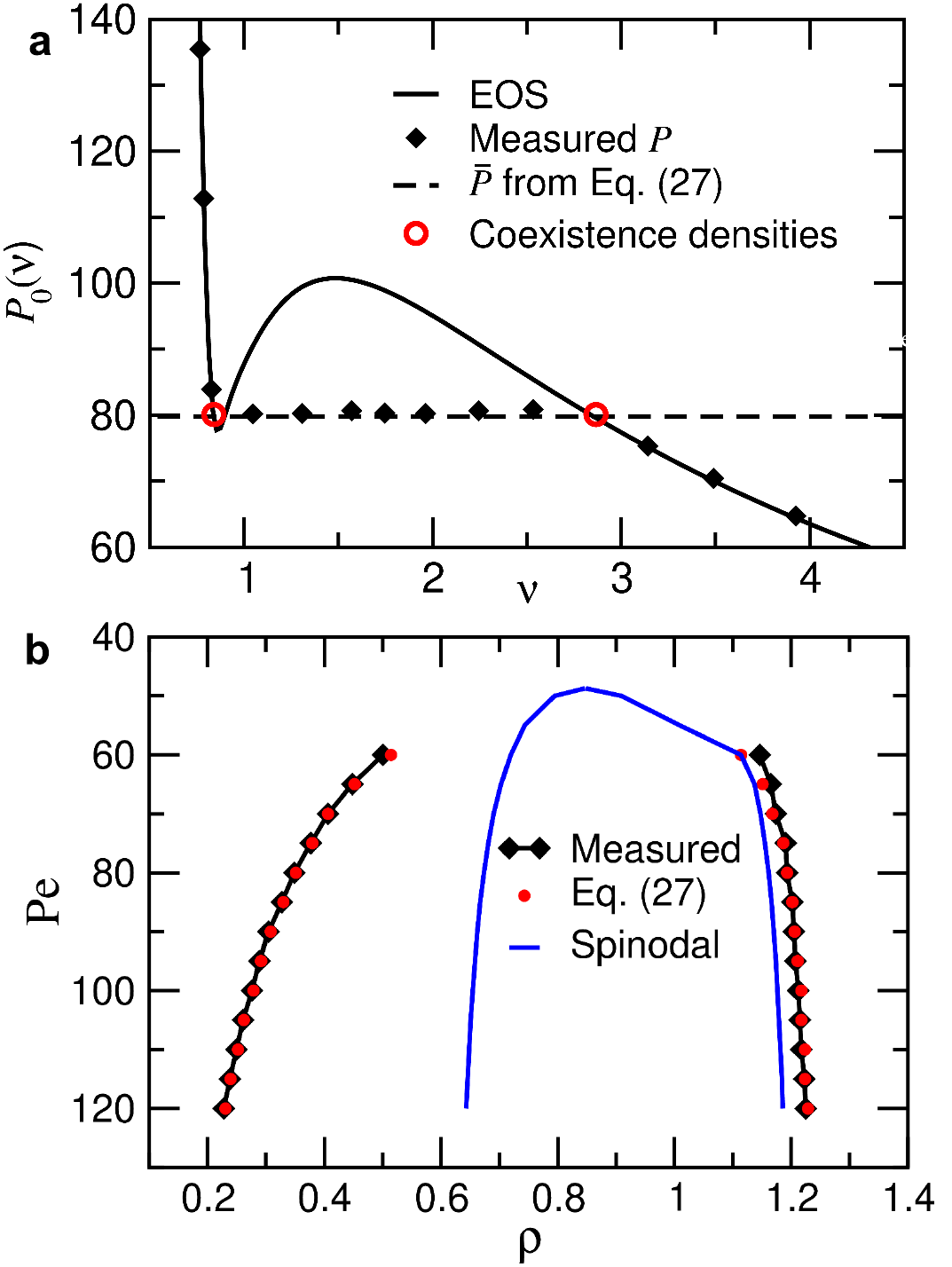}
    \caption{\textbf{``Unequal area construction'' in ABPs.} \textbf{(a)} $P_0(\nu)$ for ABPs at $\mathrm{Pe} = 80$, obtained from fitting the pressure contributions in Eq.~\eqref{P_bulk} in the low-Pe regime together with scaling arguments extending it into the two-phase region. Dashed line shows predicted coexistence pressure $\bar{P}$ according to Eq.~\eqref{MC} using values of $P_1(x)$ and $\nu(x)$ measured from simulations, diamonds the numerically measured pressures across the tieline, and red circles the measured coexistence densities, in good accordance with the theoretical prediction. \textbf{(b)} Measured (black symbols) and predicted (red symbols) coexistence densities at various Pe from Eq.~\eqref{MC}. The solid line shows the spinodal determined by the condition $dP_0/d\nu > 0$. Reproduced from~\citep{Solon:NJP:2018}.} \label{Binodals_Maxwell}
\end{figure}

\noindent An important corollary of the EOS nature of the mechanical pressure in ABPs is that the pressure in coexisting phases must be identical. For noninteracting ABPs, $P_I$ of Eq.~\eqref{P_id} is a monotonically increasing function of $\rho$, implying that phase coexistence is not possible. For interacting ABPs, the pressure is composed of one term, $P_0^D$, that increases monotonically with density (assuming that $\vect{F}$ is purely repulsive), and another, $P_0^A$, which depends on the product $v(\rho) \rho$. Since $v(\rho)$ decreases with $\rho$, $P_0^A(\rho)$ is generally non-monotonic, implying that a liquid phase at density $\rho_\ell$ and a gas phase at density $\rho_g$ can coexist at the same total pressure, $P_0(\rho_\ell) = P_0(\rho_g) = \bar{P}$ (Fig.~\ref{Pressure_EOS}b). 

To determine the coexistence densities, it is not sufficient to consider only the bulk pressure, but we rather need to analyse the full pressure profile across the interface of the coexisting gas and liquid phases. To this end, we formally decompose the total pressure $P$ as $P = P_0(\rho(x)) + P_1([\rho],x)$, where $P_1$ contains all \emph{nonlocal} contributions to the pressure, which depend on gradients of $\rho$ of all orders. Since mechanical stability requres the pressure to be constant throughout the whole system, we have that $P_0(\rho(x)) - \bar{P} = P_1([\rho],x)$ at phase coexistence. We now view the bulk pressure $P_0$ as a function of the volume per particle $\nu$ and integrate both sides to yield
\begin{equation}\label{MC}
    \int_{\nu_\ell}^{\nu_g} \left[ P_0(\nu) - \bar{P} \right] d\nu = \int_{x_\ell}^{x_g} P_1(x) \partial_x \nu dx \equiv \Delta A,
\end{equation}
where the $\partial_x \nu$ factor on the RHS comes from the variable substitution, and the integral runs from a point $x_\ell$ in the bulk liquid to a point $x_g$ in the gas, where $P_1$ vanishes. For phase separation driven by a free energy density, as described through the standard equilibrium ``Model B'' ~\citep{Hohenberg:RMP:1977}, the integrand on the RHS is proportional to the interfacial tension $\kappa$, and can be written as the total derivative~\citep{Solon:NJP:2018} 
\begin{equation}
P_1(x) \partial_x \nu = \partial_x \left( \frac{\kappa (\partial_x \rho)^2}{2\rho}\right).
\end{equation}
Since $\partial_x \rho = 0$ in both bulk phases, in the equilibrium case we get $\Delta A = 0$, so that the coexistence condition~\eqref{MC} becomes independent of $\kappa$. Equation~\eqref{MC} is then identical to the equal area construction expressed geometrically in Fig.~\ref{Maxwell}; thermodynamically, setting $\Delta A = 0$ is thus equivalent to imposing the equality of chemical potentials in coexisting bulk phases. 

For ABPs, the interfacial contributions to the pressure are given by~\citep{Solon:PRL:2015,Solon:NJP:2018}
\begin{equation}
    P_1[\rho] = P_1^D [\rho] + P_1^A [\rho] + \frac{v_0^2}{D_R}\mathcal{Q} - \frac{v_0 D_T}{D_R} \partial_x \mathcal{P},
\end{equation}
where $P_1^D$ and $P_1^A$ contains gradient contributions to the direct and active pressure that can be expressed as averages of microscopic correlators~\citep{Solon:NJP:2018}, and $\mathcal{P}$ and $\mathcal{Q}$, as before, denote the the polar and nematic order normal to the interface, both of which vanish in the disordered bulk phases. Unlike the equilibrium case, $P_1$ cannot generally be expressed as a total derivative, and we thus expect $\Delta A \neq 0$; thus, $\Delta A$ can be said to quantify the \emph{violation} of the equilibrium Maxwell construction in coexisting ABPs. If we know the EOS $P_0(\nu)$, the interfacial pressure $P_1[\nu]$, and the (inverse) density profile $\nu(x)$ for each $\mathrm{Pe}$, we can numerically calculate $\Delta A$ on the RHS, and then adjust $\bar{P}$ in the LHS integral until its value becomes equal. This value of $\bar{P}$ then corresponds to the coexistence pressure, while the corresponding $\nu_\ell$ and $\nu_g$ give the coexistence densities, in the same way as for the Maxwell construction (\emph{c.f.} Fig.~\ref{Maxwell}). An accurate form of the EOS for the full range of Pe can be constructed by fitting $v(\rho)$ and $P^D_0(\rho)$ of Eq.~\eqref{P_bulk} in the homogeneous phase ($\mathrm{Pe} < \mathrm{Pe}_c$), and extrapolating their behaviour into the $\mathrm{Pe} > \mathrm{Pe}_c$ region using straightforward scaling arguments~\citep{Solon:NJP:2018} (see Excercise 4). The interfacial terms in $P_1[\rho]$ and the density profile $\rho(x)$ can however not be straightforwardly expressed by such scaling assumptions, but can be measured from simulations in the phase-separated state set up with a flat gas-liquid interface (Fig.~\ref{Pressure_EOS}b). In this way, the binodal densities $\nu_g^{-1}$ and $\nu_\ell^{-1}$ can be predicted and compared to their directly measured values at each Pe. As seen in Fig.~\ref{Binodals_Maxwell}, this procedure confirms the validity of the criterion~\eqref{MC}, although it does not constitute a predictive theory as long as we do not have \emph{a priori} knowledge of $P_1[\rho]$.

According to~\eqref{MC}, the violation of the equlibrium equal-area construction is measured by interfacial terms in the pressure, highlighting the strong density-polarisation coupling underlying MIPS: near the gas-liquid interface, particles are on average oriented towards the dense phase, creating a net polarization. Due to the self-propulsion, this polarization is coupled to the flux balance encoded in the pressures $P_0$ and $P_1$, which in turn sets the steady-state coexistence densities. This phenomenology highlights a qualitative difference from the equilibrium case, where, as we showed above, interfacial terms in the free energy strictly do not affect properties of the bulk phases. Taken together, the results discussed here thus highlight both the suprisingly strong analogies and the subtle but important differences between MIPS and equilibrium phase coexistence at the mesoscopic scale. 

\section{Current research directions}

\noindent These notes have so far focussed on a rather narrow slice of the extensive research literature on MIPS, with an emphasis on the basic features of MIPS in QSAPs and ABPs. In this concluding section, we will give a brief overview of three of the most active research directions within the MIPS field and discuss some open problems related to these. 
\\[1mm]

\begin{figure}[h]
    \centering
    \includegraphics[width=6.5cm, clip, trim=0.0cm 0.0cm 0.0cm 0.0cm]{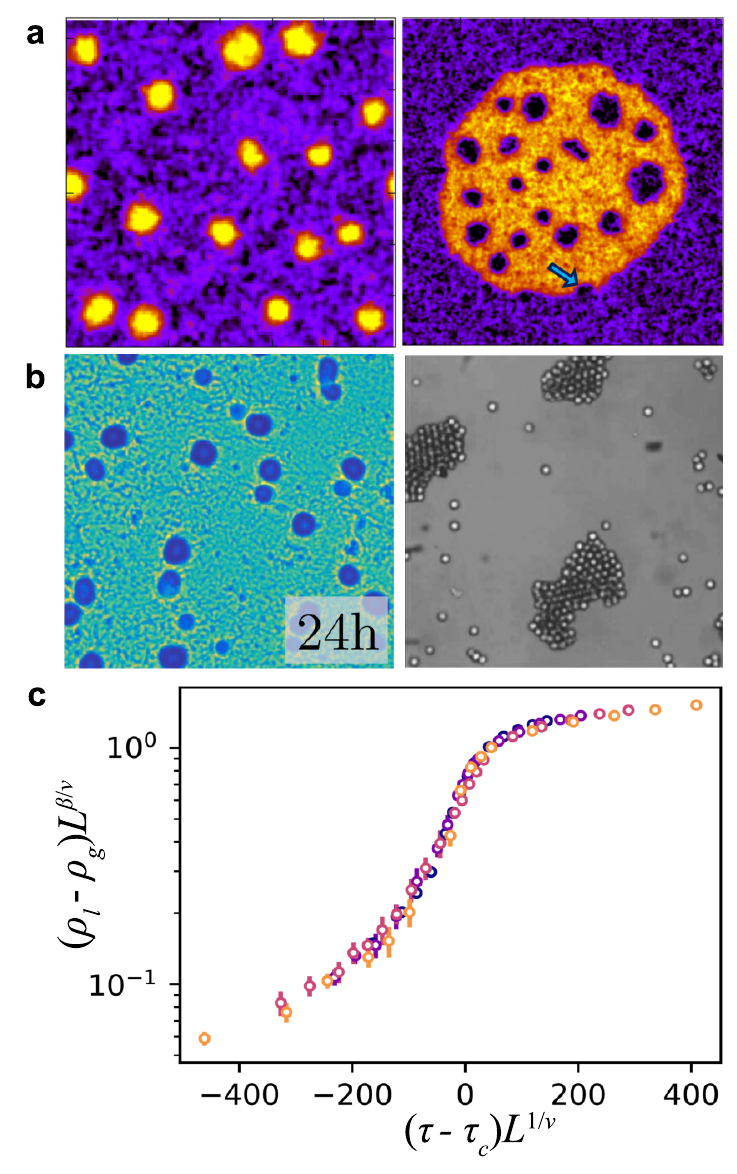}
    \caption{\textbf{Current topics of MIPS research.} \\ \textbf{(a)} Arrested (left) and ``bubbly'' (right) phase separation in Active Model B+. In both phases, detailed balance-violating terms in the current induce a reversal of the Ostwald ripening process, leading to steady states with either gas bubbles (black) or clusters (yellow) of a characteristic, finite size. Reproduced from~\citep{Tjhung:PRX:2018}. \textbf{(b)} Experimental examples of MIPS in \emph{Myxococcus xanthus} bacteria (left) and self-propelled Janus colloids (right). Reproduced respectively with permission from~\citep{Liu:PRL:2019} and~\citep{Buttinoni:2013:PRL} (\copyright 2019/2013, American Physical Society). \textbf{(c)} Critical behaviour of AOUPs. The plot shows a collapse of the difference $\rho_\ell - \rho_g$ for different system sizes $L$ (different coloured symbols) when rescaled using the Ising exponents $\nu = 1$ and $\beta = 1/8$. The abscissa shows the rescaled distance from the critical value $\tau_c$ of the AOUP persistence time $\tau$ (\emph{c.f.} Eq.~\eqref{AOUP2}). Reproduced from~\citep{Maggi:SoftMatter:2021}.} \label{Outlook}
\end{figure}

\noindent \textbf{Phenomenological field theories.} Field-theoretical descriptions of active matter models have proven one of the most powerful tools to describe their collective behaviour~\citep{Marchetti:RMP:2013,Cates:LesHouches}. Most of these are \emph{microscopic} field theories that build on an explicit coarse-graining of the equations of motion, such as those outlined in Section~\ref{sec:pressure}. These have provided, for example, predictions for the pressure~\citep{Solon:PRL:2015,Winkler:SoftMatter:2015,Speck:PRE:2016} and spinodal line~\citep{Bialke:2013:EPL,Bickmann:JPCM:2020} in terms of the microscopic parameters, often based on input data from simulations such as pair distribution functions and density profiles. An alternative approach is to use \emph{phenomenological} field theories, where equations of motion for the order parameter fields are instead formulated based on the underlying symmetries of the problem. When describing MIPS, the relevant starting point is the equilibrium Model B (or Cahn-Hilliard equation) that describes phase separation in fluids with local diffusive dynamics and conserved density~\citep{Hohenberg:RMP:1977}. This identification relies on the fact that, at bulk level, MIPS can be described by the free energy structure in Eq.~\eqref{f_0}~\citep{Tailleur:2008:PRL}. Modification of Model B to take into account the non-equilibrium nature of higher-order gradient terms and the possibility of circulating steady-state currents has respectively lead to two field theories dubbed \emph{Active Model B}~\citep{Wittkowski:2014:NC} and \emph{Active Model B+}~\citep{Tjhung:PRX:2018} that generically describe phase separation in dry, scalar active matter. For certain parameter values, Active Model B+ shows a reversal of the Ostwald ripening process, leading to arrested phase separation with finite-size particle clusters or spontaneously forming gas bubbles in the dense phase~\citep{Tjhung:PRX:2018} (Fig.~\ref{Outlook}a). The latter phenomenology -- ``bubbly phase separation'' -- is similar to what is observed in ABP simulations~\citep{Caporusso:PRL:2020,Shi:PRL:2020} (\emph{c.f.} Fig.~\ref{ABPs}), while arrested particle clustering has been observed in experiments on active Janus colloids~\citep{Theurkauff:2012:PRL,Buttinoni:2013:PRL} and simulations of ABPs with attractions~\citep{Alarcon:SoftMatter:2017}. The fundamental properties of such phenomenological field theories as well as their connection to microscopic active particle models are therefore central open questions in our theoretical understanding of MIPS. \\

\noindent \textbf{Experimental manifestations of MIPS.} Since both the QSAP, ABP, and AOUP models are idealised models of biological or synthetic microswimmers, the vast majority of results on MIPS are theoretical or computational in nature. However, the self-trapping mechanism leading to MIPS has also been confirmed in synthetic self-propelled colloids~\citep{Theurkauff:2012:PRL,Buttinoni:2013:PRL} as well as the bacterium \emph{Myxococcus xanthus}~\citep{Liu:PRL:2019} (Fig.~\ref{Outlook}b) that moves by sliding on solid substrates. However, because most experimental realisations of active particles at the microscopic level are surrounded by a fluid, the role of hydrodynamic, phoretic and other specific interactions will change the collective behaviour even though the basic MIPS mechanism remains robust. One promising avenue to quantitatively test analytical and computational predictions on MIPS is therefore to instead use macroscopic realisations of ABPs, such as grains on a vibrated plate~\citep{Deseigne:PRL:2010}, or systems of autonomous robots~\citep{Chvykov:Science:2021}. Another open problem is to build a deeper knowledge of the importance of the MIPS self-trapping mechanism in microbiology, for example in pattern formation~\citep{Cates:PNAS:2010} and the initial stages of biofilm formation~\citep{Polin:elife:2021}. \\

\noindent \textbf{Critical behaviour in MIPS.} Understanding the behaviour of phase-separating systems near their critical point is one of the major achievements of modern statistical physics. This understanding is built on the notion that systems can be grouped into a set of universality classes based on their behaviour on long length- and timescales where the microscopic details become irrelevant~\citep{Hohenberg:RMP:1977}. Since it is not \emph{a priori} obvious what universality class(es) ABPs, AOUPs and QSAPs belong to, a lot of effort has been put into determining this numerically by measuring critical exponents as the suspension approaches the MIPS critical point~\citep{Maggi:SoftMatter:2021,Partridge:PRL:2019,Siebert:PRE:2018,Dittrich:EPJE:2021}. This is highly challenging, due to the diverging correlation lengths near the critical point and the resulting strong finite-size effects. Hence, there is currently a number of conflicting findings for these exponents: Simulations of AOUP suspensions~\citep{Maggi:SoftMatter:2021} and lattice models of ABPs~\citep{Partridge:PRL:2019} point towards the same critical exponents as in the Ising universality class (Fig.~\ref{Outlook}c), just as for equilibrium gas-liquid coexistence, while other simulations of ABPs on-lattice~\citep{Siebert:PRE:2018} and off-lattice~\citep{Dittrich:EPJE:2021} show values inconsistent with the Ising exponents. Understanding these inconsistencies, and whether ABPs, AOUPs and QSAPs belong to the same universality class, thus poses a major computational and theoretical challenge. 

\section{Exercises}

\noindent \textbf{Exercise 1.} Derive Eq.~\eqref{eq:QSAP_flux} and then verify that the criterion in Eq.~\eqref{eq:vrho_const} implies $J = 0$. Start from the expressions of the density of right- and left-moving particles $R(x,t)$ and $L(x,t)$, respectively~\citep{Schnitzer:PRE:1993}:
\begin{align}
\frac{\partial R}{\partial t} &= -\frac{\partial (vR)}{\partial x} - \frac{\alpha R}{2} + \frac{\alpha L}{2}  \\
\frac{\partial L}{\partial t} &= +\frac{\partial (vL)}{\partial x} + \frac{\alpha R}{2} - \frac{\alpha L}{2} \nonumber.
\end{align}
First, combine the above equations to yield PDEs for the auxiliary fields $\rho \equiv R + L$ and $\sigma \equiv R - L$, and note that the particle current is given by $J = v \sigma$. Then, combine these equations to an expression for $\partial_t^2 \rho$ and assume that the density $\rho$ relaxes slowly enough that we can set $\partial_t^2 \rho = 0$ to yield Eq.~\eqref{eq:QSAP_flux} for $J$. See also Ref.~\citep{Schnitzer:PRE:1993}, but beware of a typo in the last term of Eq.~(2.4)! \\

\noindent \textbf{Exercise 2.} Show that Eq.~\eqref{eq:delta_rho} implies the MIPS criterion in Eq.~\eqref{MIPS_criterion}. Then, show that the latter is equivalent to a spinodal criterion ($\partial^2_\rho f_0 < 0$) on the bulk free energy density given by Eq.~\eqref{f_0}. \\

\noindent \textbf{Exercise 3.} Derive the moment equations~\eqref{m0_eqn} and~\eqref{m1_eqn} by integrating the master equation~\eqref{master_eqn}, in the latter case after multiplication with $\cos \theta$, and subsequently setting $\partial_t \rho = \partial_t \mathcal{P} = 0$. Then use the pressure definition in Eq.~\eqref{P_def} and integrate to give the expression~\eqref{P_id} for the ideal ABP pressure. For the integration limits note that, for $x = 0$ (in the disordered bulk) and $x \rightarrow \infty$ (inside the wall), we have $\mathcal{P} = \mathcal{Q} = 0$, while $\rho(0) = \rho_0$ and $\rho(x \rightarrow \infty) = 0$. \\

\noindent \textbf{Exercise 4.} An accurate EOS for interacting ABPs can be obtained by separately considering $P_0^A$ and $P_0^D$ in Eq.~\eqref{P_bulk}. Rephrased in terms of Pe, the former reads
\begin{equation}
	P_0^A = \frac{\rho v(\rho)}{6} \mathrm{Pe},
\end{equation}
where we use Lennard-Jones units for which $\sigma = 1$. The effective $v(\rho)$ for interacting ABPs is known to be linear up to high densities, where higher-order terms become important. This is captured by the following form of $v(\rho)$:
\begin{equation}
	v(\rho) = v_0 g(\rho) (1 - a_0 \rho + a_1 \rho^2),
\end{equation}
where the regularisation 
\begin{equation}
g(\rho) = \frac{1}{2}\left( 1- \tanh [a_2(\rho - \rho_*)] \right)
\end{equation}
ensures that $v$ smoothly approaches zero as $\rho$ approaches $\rho_*$, and $a_0$ -- $a_2$ are fitting parameters. 

For ABPs interacting via the WCA potential (Fig.~\eqref{fig:WCA}), $P_0^D$ is accurately described by a biexponential function:
\begin{equation}
	P_0^D = d_0 \left[ \exp(d_1 \rho) - 1 \right] + d_2 \left[ \exp(d_3 \rho) - 1 \right],
\end{equation}
where $d_1$ -- $d_3$ are fitting parameters. Note that, as long as Pe is varied by changing $D_R$ while keeping $v_0$ constant, $P_0^D$ is to a good approximation Pe-independent. 

Using the above expressions and the parameter values in Table~\ref{Parameters}, plot the EOS $P(\nu)$ with $\nu = \rho^{-1}$ for a few values of Pe below and above the transition to MIPS and approximately locate $\mathrm{Pe}_c$ by noticing where the EOS becomes nonmonotonic. Numerically perform the equilibrium Maxwell construction (Eq.~\eqref{MC} with $\Delta A = 0$) for $\mathrm{Pe} > \mathrm{Pe}_c$ and show that the predicted coexistence densities fail to reproduce the coexistence densities measured from ABP simulations given in Table~\ref{Binodal_table} and in Fig.~\ref{Binodals_Maxwell}. 

\begin{acknowledgments}
\noindent These notes are part of the “Initial Training on Theoretical Methods for Active Matter” organised by the MSCA-ITN ActiveMatter which has received funding from the EU H2020 Research and Innovation Programme under Grant Agreement No 812780. JS is funded by a project grant from the Swedish Research Council (grant number 2019-03718). 
\end{acknowledgments}

\newpage

\begin{table}[h!]
\centering
\begin{tabular}{|c|c|}
\hline
$v_0$		& $24$    	\\
\hline
$a_0$		& $0.898$    	\\
\hline
$a_1$ 		& $7.03\times 10^{-3}$   	\\
\hline
$a_2$ 		& $13.6$		\\
\hline
$\rho_*$ 	& $1.19$      \\
\hline
$d_0$  		& $4.548 \times 10^{-9}$ \\
\hline
$d_1$		& $17.79$ 	\\
\hline
$d_2$		& $0.691$	\\
\hline
$d_3$		& $3.886$ 	\\
\hline
\end{tabular}
\caption{EOS parameter values for ABPs interacting via the WCA potential~\citep{Solon:NJP:2018}.}\label{Parameters}
\end{table}

\begin{table}[h!]
\centering
\begin{tabular}{|c|c|c|}
\hline
Pe			& $\rho_g$ & $\rho_\ell$   	\\
\hline
$60$		& $0.500$ & $1.146$ \\
\hline
$70$ 		& $0.406$ & $1.174$ \\
\hline
$80$ 		& $0.349$ & $1.193$ \\
\hline
$90$ 		& $0.304$ & $1.204$ \\
\hline
$100$  		& $0.276$ & $1.212$ \\
\hline
$110$		& $0.250$ & $1.217$	\\
\hline
$120$		& $0.228$ & $1.225$	\\
\hline
\end{tabular}
\caption{Pe-dependent coexistence densities measured from ABP simulations~\citep{Solon:NJP:2018}.}\label{Binodal_table}
\end{table}

\bibliography{biblio-MIPS.bib}

\end{document}